# ARIANNA: A New Concept for UHE Neutrino Detection

**Steven W. Barwick**

Department of Physics and Astronomy, University of California, Irvine, CA 92697

E-mail: barwick@cosmic.ps.uci.edu

**Abstract**. The ARIANNA concept utilizes the Ross Ice Shelf near the coast of Antarctica to increase the sensitivity to cosmogenic neutrinos by roughly an order of magnitude when compared to the sensitivity of existing detectors and those under construction. Therefore, ARIANNA can test a wide variety of scenarios for GZK neutrino production, and probe for physics beyond the standard model by measuring the neutrino cross-section at center of momentum energies near 100 TeV. ARIANNA capitalizes on several remarkable properties of the Ross Ice Shelf: shelf ice is relatively transparent to electromagnetic radiation at radio frequencies and the water-ice boundary below the shelf creates a good mirror to reflect radio signals from neutrino interactions in any downward direction. The high sensitivity results from nearly six months of continuous operation, low energy threshold ($\sim 3 \times 10^{17}$ eV), and more than $2\pi$ of sky coverage. The baseline concept for ARIANNA consists of moderately high gain antenna stations arranged on a 100 x 100 square grid, separated by about 300m. Each station consists of a small group of cross-polarized antennas residing just beneath the snow surface and facing downwards. They communicate with a central control hub by wireless links to generate global triggers.

## 1. Introduction

The scientific promise of high energy neutrino astronomy remains as compelling and elusive as ever. Although powerful neutrino telescopes such as AMANDA-II [1] and NT-200 in Lake Baikal [2] have uncovered no evidence for astrophysical neutrino sources, these first-generation detectors, optimized to detect neutrinos with energies between $10^{12}$-$10^{15}$ eV, have paved the way for more capable telescopes with instrumented volumes as large as one cubic kilometer [3]. At yet higher neutrino energies, new techniques were developed that utilize *coherent* Cherenkov emission at *radio* wavelengths. This emission mechanism, known as the Askaryan effect [4], was experimentally confirmed [5] less than a decade ago. The balloon-borne ANITA-lite [6] payload and the South Pole based RICE [7] array have exploited this effect in Antarctic ice to produce important constraints on the extraterrestrial neutrino flux. In December 2006, the fully-instrumented ANITA [8] payload is scheduled to launch from McMurdo, Antarctica. Since the power of coherent radio emission grows as the square of the shower energy (and therefore neutrino energy), the balloon-borne detectors tend to yield interesting apertures above $10^{18}$ eV. Therefore, a gap exists in the energy coverage of current-generation high energy neutrino detectors, as shown in figure 1. ARIANNA, first proposed [9, 10] about one year ago, is designed to bridge this gap in sensitivity. The international "concept development team" includes physicists from UC-Irvine, UC-Los Angeles, Univ. of Hawaii, Ohio State Univ., and Univ. College London. Team members gained extensive experience by participating in ongoing Antarctic neutrino programs (AMANDA-II, IceCube, ANITA) and an a large area, remotely deployed, air shower array located in Argentina (Pierre Auger Observatory).

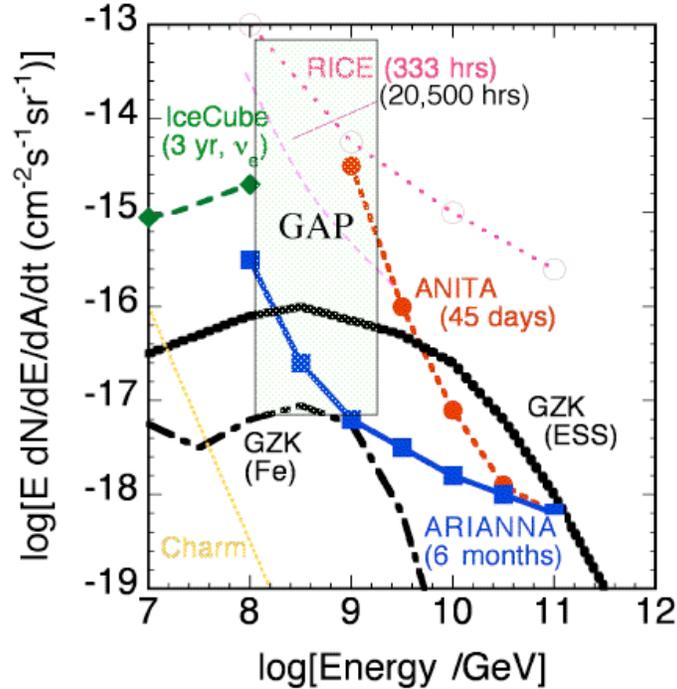

**Figure 1**. Representative survey of experimental flux limits, anticipated sensitivity of current generation instruments, and theoretical predictions for neutrino energies in excess of $10^{16}$ eV. Note the gap in instrumental sensitivity between $10^{17}$ eV and $\sim 3 \times 10^{18}$ eV. ARIANNA is designed to provide sufficient sensitivity to bridge the energy gap. Auger (not shown) [11], which detects tau neutrinos, is approximately a factor of 10 less sensitive than ARIANNA.

1. **Scientific rationale for ARIANNA**

Cosmic rays are known to extend to energies above $10^{20}$ eV, and the very highest of those almost certainly originate outside the galaxy. Griessen, Zatsepin, and Kuzmin[12] (GZK) first recognized that cosmic rays with energies in excess of $3 \times 10^{19}$ eV readily interact with cosmic microwave photons and lose energy quickly, thereby limiting their propagation distance to the local supercluster. To date the measured energy spectrum shows hints, but no compelling feature, due to the GZK absorption. Even more confounding, no known sources are known to exist within this volume. This leads to uncertainty in both the nature of the sources and their spatial distribution. Recently, Seckel and Stanev[13] have pointed out that neutrinos may be the key to unraveling the mysteries of the highest energy cosmic rays. For example, the neutrino energy spectrum helps to break the degeneracy in the model dependence between source distribution and evolution.

Neutrinos spanning the energy interval $10^{17-20}$ eV are produced as a direct by-product the GZK mechanism[14]. Since the GZK mechanism is based on widely accepted tenets and measurements, the predictions for the neutrino flux is perhaps the most secure of any in neutrino astronomy. GZK neutrinos collide with matter on earth with center of momentum energies $\sim$100 TeV, and thereby provide an opportunity to study physics at energies above that which is available at current or planned accelerator facilities. ANITA, expected to launch in December 2006, has sufficient sensitivity to make the first detection of GZK neutrinos. In this paper, I describe a new concept for a detector with an order of magnitude increase in collecting power, which is called ARIANNA (the Antarctic Ross

Iceshelf ANtenna Neutrino Array). ARIANNA exploits an unprecedented opportunity to probe the energy frontier of particle physics with a "beam" that interacts solely by the weak force.

Several ideas exist in the literature [15] to measure the neutrino cross-section at extremely high energy. The method of Kusenko and Weiler combine information gathered by horizontal air shower detectors and extensive air shower arrays. Their method depends on the reliable assessment of common systematic errors for different detectors. The method outlined by Anchordoqui, et al., compare the rate of earth skimming neutrinos to the rate of downward traveling neutrinos in buried optical arrays such as AMANDA and IceCube. This technique depends on a precise knowledge of the angular distribution of signal and background for all neutrino flavors. Granted sufficient statistics, ANITA may be able to measure the neutrino cross-section [8] by counting the number of neutrino events from the *ice sheet* and *ice shelf* regions. The event rate in the ice shelf depends linearly on the cross-section whereas the rate from the ice sheet depends relatively weakly on the cross-section. The utility of the technique depends on the actual flux; higher flux provides access to smaller deviations from Standard Model predictions. Although the GZK neutrino flux is almost guaranteed [16], it is very small. ARIANNA has sufficient collecting power to ensure adequate statistics to determine the cross-section from the zenith angle dependence of the measured flux.

The scientific advantages of ARIANNA are summarized as follows:

ARIANNA increases the sensitivity for the detection of GZK neutrinos by an order of magnitude over state-of-the-art detectors currently under construction, such as ANITA. Simulations indicate that ARIANNA can observe **~40 events per 6 months** of operation based on the widely-used predictions for the GZK neutrino flux by Engel, Seckel and Stanev (ESS) [16] .

The low energy threshold of ARIANNA, combined with high statistics, provides an unparalleled opportunity to measure a broad interval of the neutrino energy spectrum. Since the receiver antennas are relatively close to the source of the signal, the energy threshold is correspondingly lowered. Simulations show that the fraction energy resolution is $\Delta E/E \sim 1$.

ARIANNA can test alternative scenarios for GZK neutrino production. For example, models [17] that assume that extragalactic cosmic rays are mixed elemental composition, or perhaps entirely iron nuclei, predict fluxes that may be as small as ~5% of the ESS predictions.

ARIANNA can survey the southern half the sky for point sources of high-energy neutrinos with unprecedented sensitivity. Preliminary reconstruction studies (see figure 2) show that the neutrino direction be measured to a precision of 1 degree, and additional improvements in the reconstruction procedure are expected. These encouraging results show that good reconstruction can be achieved despite imprecise knowledge of the path of the signal.

ARIANNA can probe for physics beyond the standard model by measuring the neutrino cross-section at center of momentum energies of 100 TeV, a factor 10 larger than available at the LHC. Figure 3 shows the angular distribution of reconstructed events if the cross-section is standard model, and twice and half standard model. The two-parameter fit of the reconstructed data involves normalization and cross-section. Preliminary studies [18] indicate that the cross-section can be measured with a precision of 25%, benefiting from the large statistical sample of 400 events spanning $2\pi$ solid angle.

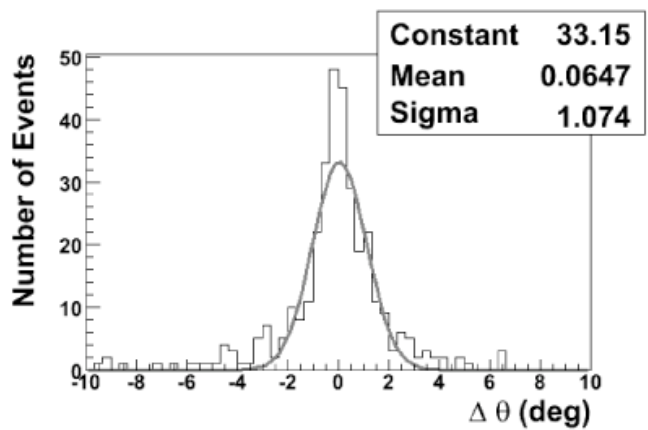 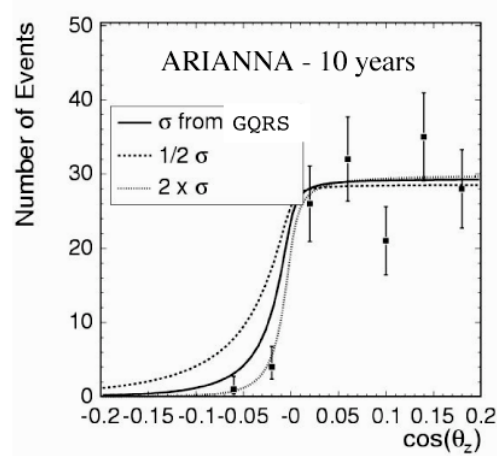

**Figure 2:** Preliminary estimate of angular resolution of ARIANNA for 4-station trigger. Event reconstruction includes timing and amplitude information from each station, but does not use the polarization measurements. Not surprisingly, studies show that angular resolution improves as the number of stations increases. Good reconstruction is achieved despite the ambiguity in the trajectory of the conical radio pulse.

**Figure 3.** Comparison of expected zenith angle distribution of neutrino events (zoomed to near the horizon) with the reconstructed zenith angles (filled squares) assuming 3 values of neutrino cross-section (standard model = $\sigma_{GQRS}$) [19] and isotropic source of high energy neutrinos with GZK energy spectrum.

## 2. Description of ARIANNA Baseline Design

ARIANNA achieves its unprecedented sensitivity by capitalizing on several remarkable properties of the Ross Ice Shelf: (1) shelf ice is relatively transparent to electromagnetic radiation at radio frequencies and, (2) the water-ice boundary creates a good mirror to reflect radio signals from neutrinos propagating in any downward direction relative to the ice surface (see figure 4). Therefore, ARIANNA can survey more than half the sky for point and diffusely distributed sources of ultra-high energy neutrinos. Its unprecedented sensitivity to GZK neutrinos is a consequence of nearly six months of continuous operation, low energy threshold (~$3 \times 10^{17}$ eV), and more than $2\pi$ of sky coverage.

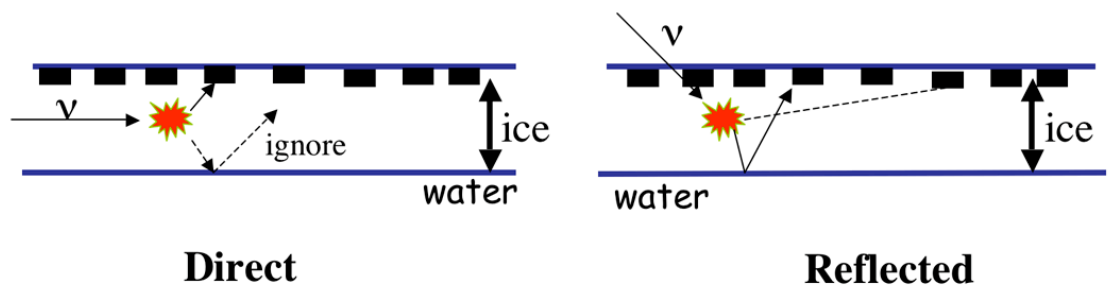

**Figure 4.** Schematic representation of "direct" (left) events that originate in the ice shelf and "reflected" (right) events. Direct signals are more important for neutrino directions that are nearly parallel to the ice-air boundary. The solid arrow represents the fraction of the Cherenkov cone that produces the largest signals in the receiver stations (black rectangles). The dashed component of the conical emission is more strongly attenuated due to the longer path length, and ignored. Downward neutrinos (right panel) are detected by reflected rays only.

ARIANNA is designed to be relatively insensitive to the most imprecisely known ice properties, such as localized patches of poor ice or a non-specular component of reflection. It does not depend significantly on the surface conditions and snow topology at the site. Station deployment is mechanically straightforward, requiring only a backhoe to dig a trench for the log-periodic receiver antennas and a shallow pit on the snow surface for the electronics module. The close proximity of the site to McMurdo should ameliorate the logistical stress on support infrastructure.

2.1. Description of ARIANNA Architecture

The baseline concept for ARIANNA consists of 100 x 100 antenna stations arranged in a square grid with lattice separation of ~300 meters. The complete array occupies an area of 30 km by 30 km, and located on the surface of the Ross Ice Shelf. Each station contains two to five radio antenna of modest gain with dual polarization capabilities and face downward. Stations may also contain antenna with improved side viewing capabilities. We expect to bury the antenna and signal processing electronics just under the snow surface to minimize temperature fluctuation and reduce the impact on the local environment. Due to the proximity to the surface, each station can be routinely serviced or repaired. Signal processing at each station relies on experience with ANITA and the Pierre Auger Observatory (PAO) to develop autonomous, reliable electronics packages. We envision that solar cells will provide the required power (~10W) to each station, extending operations to about half the year. As the sun sinks below the horizon, each station will perform "safing" procedures and await sunlight to begin operation the following year. A sealed-gel battery provides station power during short periods when clouds or fog occults the sun. At the termination of the project, ARIANNA stations can be removed from Antarctica to minimize the long-term environmental impact.

Low power communication links will transmit local trigger information to neighboring stations and to a centralized hub. GPS provides station location, and more importantly, time stamps with an accuracy of ~25 ns, which exceeds the requirements for good reconstruction. If warranted, the timing accuracy can be improved by relatively standard techniques. The system architecture of ARIANNA also borrows strongly from experience with the Pierre Auger Observatory. Local coincidence requirements at each antenna station and those within a short distance of a given station will reduce the trigger rate generated by thermally induced random noise to ~10 Hz. Each station communicates local trigger information with a centralized control hub and to neighboring stations. A master trigger is formed at the control hub and retransmitted back to the antenna stations, which then transmit the waveform data to the control hub for data archiving.

We have identified two potential sites, near Minna Bluff, based on the following criteria:

Sufficient depth of the ice shelf: Bedmap [20] data show that the Ross Ice Shelf is almost 500 (400) meters thick at the site just south (east) of Minna Bluff.

Good specular reflection: previous measurement of reflection properties [25] is encouraging.

Geographical proximity to logistical support: both sites are within helicopter range of McMurdo, and they are close to the routes developed for the Antarctic Traverse.

Expectation of low levels of anthropogenic radio noise: The site just south of Minna Bluff offers protection from Radio Frequency Interference (RFI) generated at McMurdo Station and Black Island, a communications outpost. Moreover, inevitable growth of communications traffic in the vicinity of McMurdo is unlikely to adversely affect ARIANNA.

Region free from heavy crevassing:

We expect to work closely with the National Science Foundation and its support agencies to identify the best site for ARIANNA. For example, it is desirable to find a site that minimizes the variation in the vertical ice density. In addition, we expect to utilize the wealth of experience with solar power that have been developed for instrumentation.

2.2. ARIANNA simulation and predicted capabilities

ARIANNA simulation tools were developed from two relatively mature simulation packages utilized by the ANITA collaboration. Radio emission is estimated via standard parameterizations that were validated at accelerators. For concreteness, each station consists of two dual-polarization antennas with gain and angular sensitivity identical to the measured properties of the Seavey antenna employed by ANITA. The beam pattern of a single Seavey is assumed to represent approximately the pattern of two planar log- periodic antennas oriented at right angles. The antennas are embedded in the topmost meter of the ice shelf, with a density of 0.6 g/cm$^3$ (i.e., we assume a small variation in the index of refraction due to surface snow or firn ice) and face downward. The shelf is assumed to be 500 meters thick, governed by a depth-dependent temperature profile [21]. The attenuation length for radio propagation depends on both frequency and initial depth of interaction. The modeled frequency dependence is compatible with theory [22] and recent measurements obtained at the South Pole [23]. The simulations account for the warmer and thinner ice in the Ross Ice Shelf. Averaged over the vertical temperature profile, the effective power attenuation length, $\lambda_{eff}$, for one-way radio propagation through the ice-shelf is 200m, which is consistent with values used in previous studies of the Ross Ice Shelf [24]. The reflected radio power from the seawater - ice boundary beneath the Ice Shelves is quite large. Neal [25] shows that the reflection loss is typically less than -3dB over a majority of the Ross Ice Shelf, and this value is assumed in the baseline simulations. We assume that the reflected Askaryan signal maintains phase coherence, and ignore the effect of surface roughness. Neal [26] argues that large values for reflection efficiency are correlated with a relatively small amount of surface roughness at the saltwater-ice interface, and he shows that the variation in depth from a smooth surface is 0.03m in a region of the Ross Ice Shelf that exhibits low reflection losses. Recently, Nicholls *et al*. [27] produced a sonar map of the water-ice surface beneath the Fimbul Ice Shelf. They corroborate that the rms variation of surface height is small: finding it to be less than 0.008m in regions outside of flow traces.

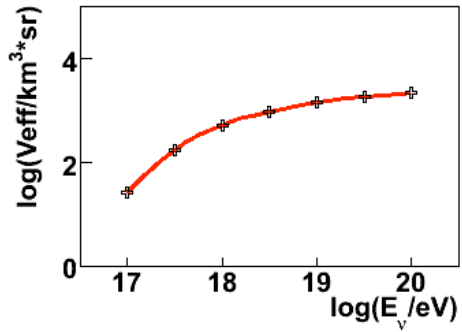 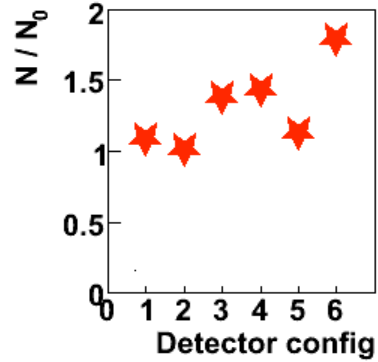

**Figure 5.** Aperture of ARIANNA (baseline configuration 2) as a function of neutrino energy.

**Figure 6.** Relative GZK event rates for six different configurations. See table 1 for details.

The simulations in this study account for propagation of neutrinos through the earth crust models and standard atmosphere, for both charged-current and neutral current interactions of all neutrino flavors, for inelasticity, for LPM effects. Secondary interactions by the outgoing charged leptons are also included. A local trigger is formed at a given antenna station when the detected power exceeds a specified threshold. Typically, we require that the signal in 2 of the 4 polarization channels exceed 2.3$V_{rms}$, where $V_{rms}$ is the rms noise generated by thermal fluctuations of ice with a vertically averaged temperature of -10C. A master trigger is generated when a minimum of 3 stations report a local trigger within a 20 microsecond time window. The arrival time is determined at each station by local

GPS clocks with a precision of 25 ns. The energy and direction of events that satisfy the trigger requirements are reconstructed by maximum likelihood methods. Since there is no easy way to tell if the radio Cherenkov signal propagated directly to the receiver antenna or reflected from the bottom, both trajectories are considered for each station. These initial studies demonstrate that good reconstruction (see figure 3) can be achieved despite the ambiguity of signal path.

Figure 5 shows that the aperture of ARIANNA becomes significant for neutrino energies between $10^{17}$ eV and $10^{18}$ eV for the baseline design (given by Config=2 in both figure 6 and table 1). The aperture, or effective volume, $V_{eff}$, integrates over solid angle and includes attenuation by the earth. Using the results of figure 5, ARIANNA will detect ~40 GZK events per calendar year. Figure 6 summarizes the results of varying the antenna configuration and local trigger requirements. The main conclusion from these studies is that (1) the baseline design is conservative and (2) it is hard to do much better than the baseline by varying the configuration. Configuration 3 and 6 achieve their modest increase by employing unrealistic trigger conditions, and 4 employs three more antennas per station.

**Table 1.** Description of ARIANNA configurations used in simulation. Unless otherwise noted, all triggers require a coincidence of 3 stations and channel threshold is set to 2.3$V_{rms}$. The configurations with 5 Quad-ridge antenna are arranged with four antenna on the corners of a square and one in the center. The "spread" dimension gives the length of one side of the square.

| Config. | Description |
|---|---|
| 1 | 1 Quad-ridge dual polarization: 1 of 2 channels |
| 2 | 2 Quad-ridge dual polarization: 2 of 4 channels  - BASELINE |
| 3 | 1 Quad-ridge dual polarization, 1 dipole: 1 of 3 channels |
| 4 | 5 Quad-ridge dual polarization, closed packed:  2 of 10 channels |
| 5 | 5 Quad-ridge dual polarization, spread by 5 m:  2 of 10 channels |
| 6 | 5 Quad-ridge dual polarization, spread by 5 m:  2 of 10 channels, 1 station coincidence |

## 3. Future Plans

With the pioneering development of TeV-scale neutrino facilities, such as AMANDA-II and NT-200 in Lake Baikal, the requisite tools to inaugurate multi-messenger astronomy already exist. IceCube, currently under construction, is designed to search for TeV-PeV neutrinos with greater sensitivity. To probe neutrino fluxes at higher energies, new techniques are being developed based on coherent radio Cherenkov emission from neutrino-induced particle cascades. ANITA holds great promise to observe the long-sought GZK neutrinos because it extends the search volume to about one million cubic kilometers. The ARIANNA concept presented at this workshop is designed to bridge the gap in sensitivity between buried cubic kilometer optical arrays and ANITA, although it is far from the only contender in the race to develop more powerful, lower-energy threshold detectors based on Askaryan's ideas [28]. We plan to investigate the capabilities of ARIANNA to identify neutrino flavor. It may be possible to distinguish nearly horizontal $\nu_\tau$ via the double bang mechanism, and $\nu_\mu$ from the "fuzz" of secondary particle cascades induced by the outgoing high-energy muon.

Obviously, the ARIANNA concept relies on the idea that suitable sites on the Ross Ice Shelf can be identified which exhibit sufficient transmission and reflection characteristics. A proposal to measure the background noise, and the attenuation and reflection properties between 100 MHz and 1 GHz in a region of ice just south of Minna Bluff (about 150 km from McMurdo Station) was approved by NSF. These tests, which start in November 2006, are expected to provide conclusive data on the viability of ARIANNA. Once established, we plan to deploy a 200-station sub-array to confirm the sensitivity estimates and evaluate the most cost-effective technologies. Work has already started on this goal. Earlier this year, the ARIANNA team completed construction and testing of the first ARIANNA prototype station, and hope to assess the robustness and reliability of the technology while deployed on the Ice Shelf.

**Acknowledgements**

The author would like to acknowledge support from the National Science Foundation (ANT-0609489), NASA (NAG5-5388), and COR funding from the University of California – Irvine. In addition, the organizers of the workshop deserve our gratitude for their tireless effort on behalf of the participants.